\documentclass[12pt,preprint]{aastex}

\shorttitle{Chemical Inhomogeneties in NGC~6791}
\shortauthors{Carrera}

\begin{document}


\title{Analysis of the CN and CH molecular band strengths in stars of the open cluster NGC~6791}


\author{R. Carrera}
\affil{Instituto de Astrof\'{\i}sica de Canarias, Spain}
\affil{Departamento de Astrof\'{\i}sica, Universidad de La Laguna, Spain}
\email{rcarrera@iac.es}

\begin{abstract}

Low resolution SDSS/SEGUE spectra have been used to study the behavior of the strengths of the CN and 
CH molecular bands in stars at different evolutionary stages of the open cluster NGC~6791.
We find a significant spread in the strengths of the CN bands, more than twice
that expected from the uncertainties, although the  
bimodalities observed in globular clusters 
are not clearly observed here. This behavior, is observed  not only
among red clump objects but also in unevolved stars such as those in the main
sequence and lower red giant branch. In contrast, not all the stars studied  show significant scatter in their CH strengths.

\end{abstract}



\section{Introduction}\label{introduction}

From the pioneering work of \citet{osborn1971}, variations in the strength of CN
molecular bands have been observed in globular cluster stars of similar
luminosities at different evolutionary stages
\citep[e.g.][]{cannon1998,cohen1999,harbeck2003}. The strength of these
molecular bands depends on temperature, surface gravity, and chemical
composition. Therefore, the fact that stars of similar luminosity and at the
same evolutionary stage have different CN band strengths denotes a different
chemical composition. These variations in  CN band strength are often
anti-correlated with the CH band intensity at
4300\AA~\citep[e.g.][]{cannon1998,harbeck2003}. Moreover, detailed analyses
based on high-resolution spectroscopy reveal that  variations in CN strengths
are accompanied by the existence of anticorrelations of the chemical abundances
of other light elements (Na, O, Al, and Mg), but not for heavier elements such as
Fe or Ca \citep{carretta2009a,carretta2009b}.

Several hypotheses concerning intrinsic and extrinsic factors have been proposed to
explain these results \citep[see][for a recent review]{pancino2010b}. However,
the detection of multiple evolutionary sequences in almost all clusters properly
studied \citep[e.g.][]{piotto2007} indicates that the most promising one is the
so-called self-enrichment scenario. First proposed by \citet{dacosta1980}, this hypothesis
assumes the existence of two or more subsequent stellar generations. The first
stars formed pollute the gas from which the second population is formed. It
seems that almost all globular clusters are massive enough to host at least two
stellar populations \citep[e.g.][]{carretta2010}. It is therefore important to investigate what happens in less
massive systems such as  open clusters, which are usually  younger,
less massive and more metal-rich than globulars. However, studies based on open clusters using both the
strength of CN bands in low-resolution spectra \citep{norris1985,hufnagel1995,martell2009} and
chemical abundances derived from high-resolution spectra
\citep{desilva2009,smiljanic2009,pancino2010a,carrera2011} have not found the same
trends observed in globulars.

There is an intriguing stellar system, NGC~6791, with a mass intermediate between those of open and
globular clusters. NGC~6791 is an ideal target for exploring the existence of chemical
inhomogeneities in its stellar population. It is traditionally
cataloged as an open cluster. However, its high mass
\citep[$5\times10^4$M$_{\odot}$][]{platais2011}, old age \citep[$\sim$ 8 Gyr;
e.g.][]{carraro2006}, and high metal content \citep[$\sim$+0.4
dex][]{carraro2006,origlia2006,carretta2007} differentiate this system 
from other open clusters. Furthermore, it is located about 1 kpc above the Galactic
plane and the eccentricity of its orbit is greater than the typical value for 
old open clusters \citep{bedin2006}. Previous determinations of NGC~6791's orbit
are compatible with an extragalactic origin \citep{carraro2006}.
However, proper motions derived from  \textit{Hubble Space Telescope} observations
suggest that this cluster was formed near the Galactic bulge with a higher initial
mass \citep{bedin2006}. The total mass of NGC~6791 has  decreased over time owing to
repeated crossings of the denser parts of the disk. The color--magnitude diagram of NGC~
6791 shows an unexpectedly wide red giant branch (RGB) and main-sequence (MS) in
the region of the turn-off. Some authors have explained these features in terms of the
existence of differential line-of-sight reddening 
\citep{platais2011}. An alternative explanation, however, is the existence of an
age spread of about 1 Gyr \citep{twarog2011}. As well as its red clump (RC),
typical of a metal-rich population, NGC~6791 also has a handful of extreme blue
horizontal branch stars whose origin is still unclear
\citep{platais2011,buzzoni2012}. \citet{kalirai2007}
associate these with strong winds due to the high-metallicity at the end of the
RGB, while \citet{bedin2008} suggest that they are the consequence of a high
binary fraction. Recent investigations of globular clusters associate these
extreme blue horizontal branch stars with the existence of a He-rich population
\citep[e.g.][]{gratton2010} although this issue remains unsolved.
\citet{janes1984}, using DDO photometry, found no signs of variations in the
compositions of the RGB stars analyzed. In contrast, \citet{hufnagel1995} found
strong indications of intrinsic differences among the CN molecular band
strengths in the RC stars of NGC~6791. Investigations of the
composition of the stars in NGC~6791 based on high-resolution spectroscopy have
found no evidence of chemical inhomogeneities \citep{origlia2006,carretta2007},
although an unexpected scatter in the Na abundance has been reported by
\citet{carretta2007}.  \citet{geisler2012} were recently the first to report the
existence of anticorrelations in the  Na and O abundances in  NGC~6791, similar
to those observed in globular cluster stars. The goal of this paper is to investigate the behavior of the
CN and CH molecular band strengths  in the stars of NGC~6791  at different
evolutionary stages. 

This paper is organized as follows: the
observational material is described in Section~\ref{observationaldata};
molecular indexes are defined in Section~\ref{indexdefinition}; the CN and CH
distributions are obtained and analyzed in Section~\ref{CNCHdistributions}; and the
main results of this work are discussed and summarized in
Section~\ref{discussion} and Section~\ref{summary}, respectively.

\section{Observational Material}\label{observationaldata}

\begin{figure}
\includegraphics[scale=0.8]{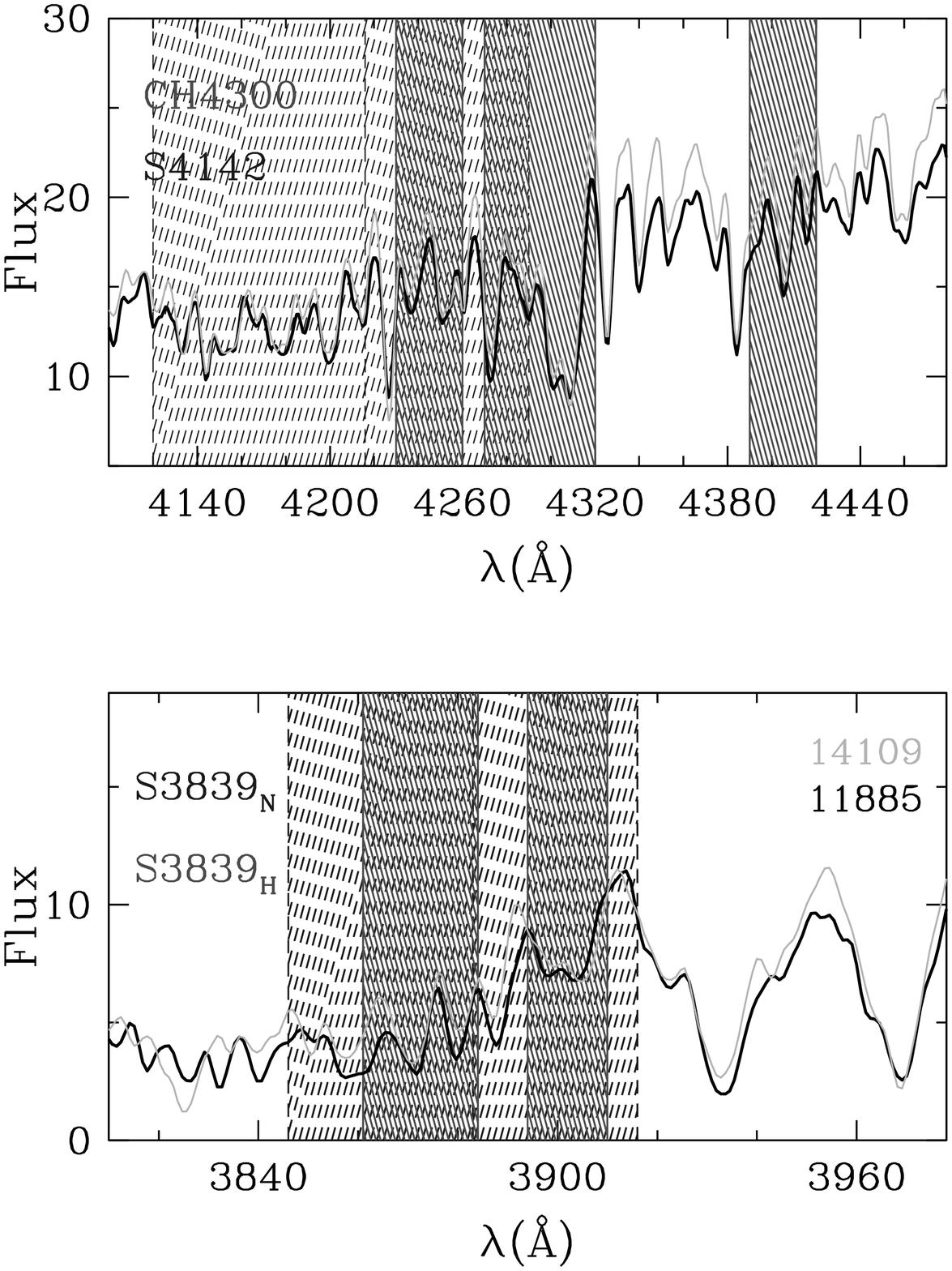}
\caption{Example of two CN-weak (gray line) and CN-strong (black line) spectra
for two stars in the lower RGB. The IDs of these stars in the \citet{stetson2003}
reference system are 14109 and 11885, respectively. The windows used for each
index have been overplotted.\label{fig_spectra}}
\end{figure}

In the framework of the Sloan Extension for Galactic Understanding and
Exploration (SEGUE)
survey \citep{yanny2009}, low resolution ($R \sim 2000$) spectra in the wavelength
range of 3600--9200\AA~were obtained for more 
than 1000 stars observed in two plates in the line of sight of NGC~6791 . They were
observed to validate the SEGUE
Stellar Parameter Pipeline \citep[SSPP,][]{allendeprieto2008,lee2008a,lee2008b}.
The reduced spectra, together with the 
atmosphere parameters, radial velocities, etc., derived for observed stars
were made public in the eighth Sloan Digital Sky Survey (SDSS) data release
\citep{aihara2011}. The procedure for obtaining 
this information can be summarized as follows. Raw spectra are first reduced by
the SDSS spectroscopic reduction pipeline, described in detail by
\citet{stoughton2002}, which provides flux- and 
wavelength-calibrated spectra, together with initial determinations of the
radial velocities and spectral types. More accurate radial velocities are
calculated in a subsequent step by SSPP, together with 
determinations of metallicity, effective temperature, and surface gravity. Two
examples of the spectra used in this paper are shown in Figure~\ref{fig_spectra}.

\begin{figure}
\includegraphics[scale=0.7]{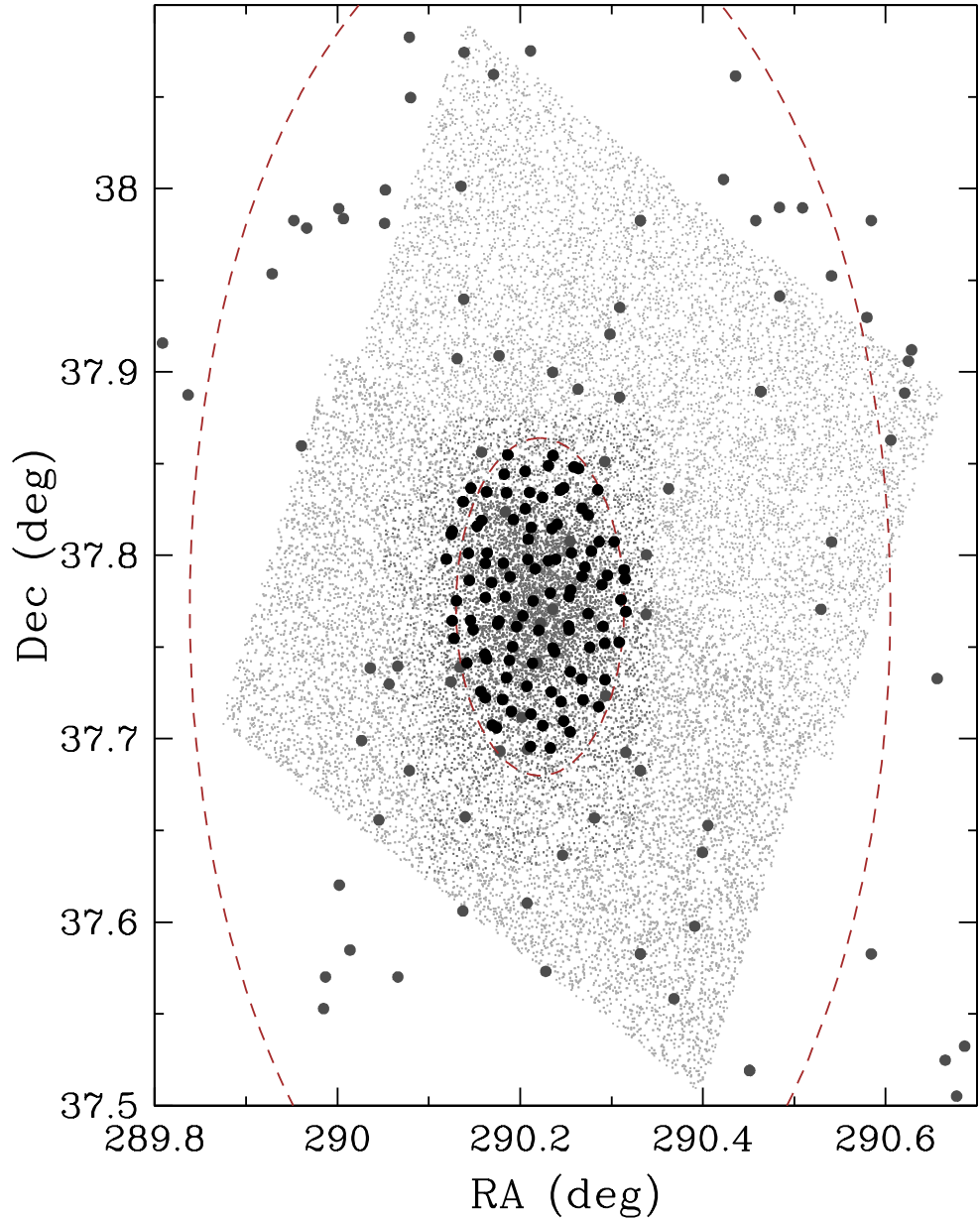}
\caption{Stars with available photometry obtained by \citet[][light gray
points]{an2008} and \citet[][dark gray points]{stetson2003} in the NGC~6791
region. Stars observed spectroscopically by SEGUE and selected as NGC~6791
members have been plotted as black circles. Gray circles denote SEGUE stars
discarded as cluster members. The dotted line denotes the tidal radius while the dashed
line shows the radius used here to select NGC~6791 members (see text for
details).\label{fig_spatial_distribution}}
\end{figure}

\begin{figure}
\includegraphics[scale=0.8]{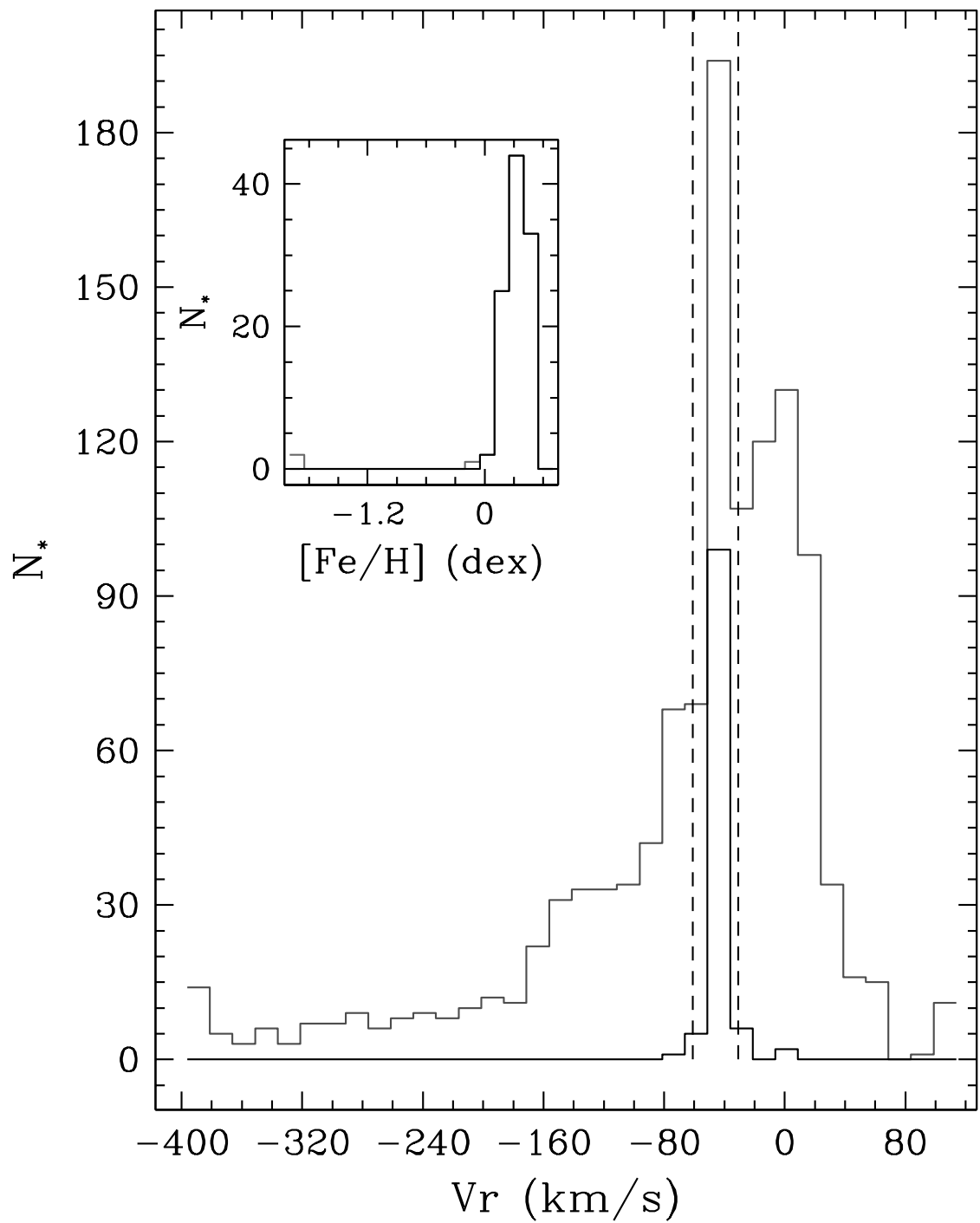}
\caption{Radial velocity histograms of all  stars observed by SEGUE (gray) and
those located within  5\farcm5 (black). Dashed lines denoted the radial velocity
limits used. [Fe/H] distributions of  stars selected from their radial velocity
and from their [Fe/H] have been plotted in  the inset panel with gray and black lines, respectively.\label{fig_distributions}}
\end{figure}

To select members of NGC~6791 from all the observed stars we followed a similar
procedure as that described by \citet{smolinski2011a}. We first reject those
stars located farther than 5\farcm5\footnote{This
 value is $1.25\times r_h$ and $0.25\times r_t$ where $r_h=$4\farcm4 and
$r_t=$23\farcm, respectively \citep{platais2011}.} from the cluster
center\footnote{$\alpha_{2000}=19^{\rm h}20^{\rm m}53^{\rm s}$ 
$\delta_{2000}=+37\arcdeg46\farcm3$} 
(Fig.~\ref{fig_spatial_distribution}). This radius
was chosen to ensure that  stars selected were among those observed in the two photometric
samples used throughout this paper and described 
below. We then discard those stars whose radial velocities
(Fig.~\ref{fig_distributions}) are not within the range $\langle
V_r\rangle_{\rm NGC\,6791}\pm 1\sigma$, assuming $\langle
V_r\rangle_{\rm NGC\,6791}$=-46$\pm$10
km s$^{-1}$ \citep{carrera2007}. In the same way, we reject those stars whose
metallicities (Fig.~\ref{fig_distributions}) do not satisfy [Fe/H]$\pm 2 \sigma$
with
[Fe/H] = +0.3$\pm$0.15 dex \citep[][and references therein]{carrera2011}. A total
of 104 stars were selected as NGC~6791 members according to these criteria.
Although we used the same data as \citet{smolinski2011a}, our sample is bigger because we use a larger radius to select the cluster members. 

\begin{figure}
\includegraphics[scale=0.7]{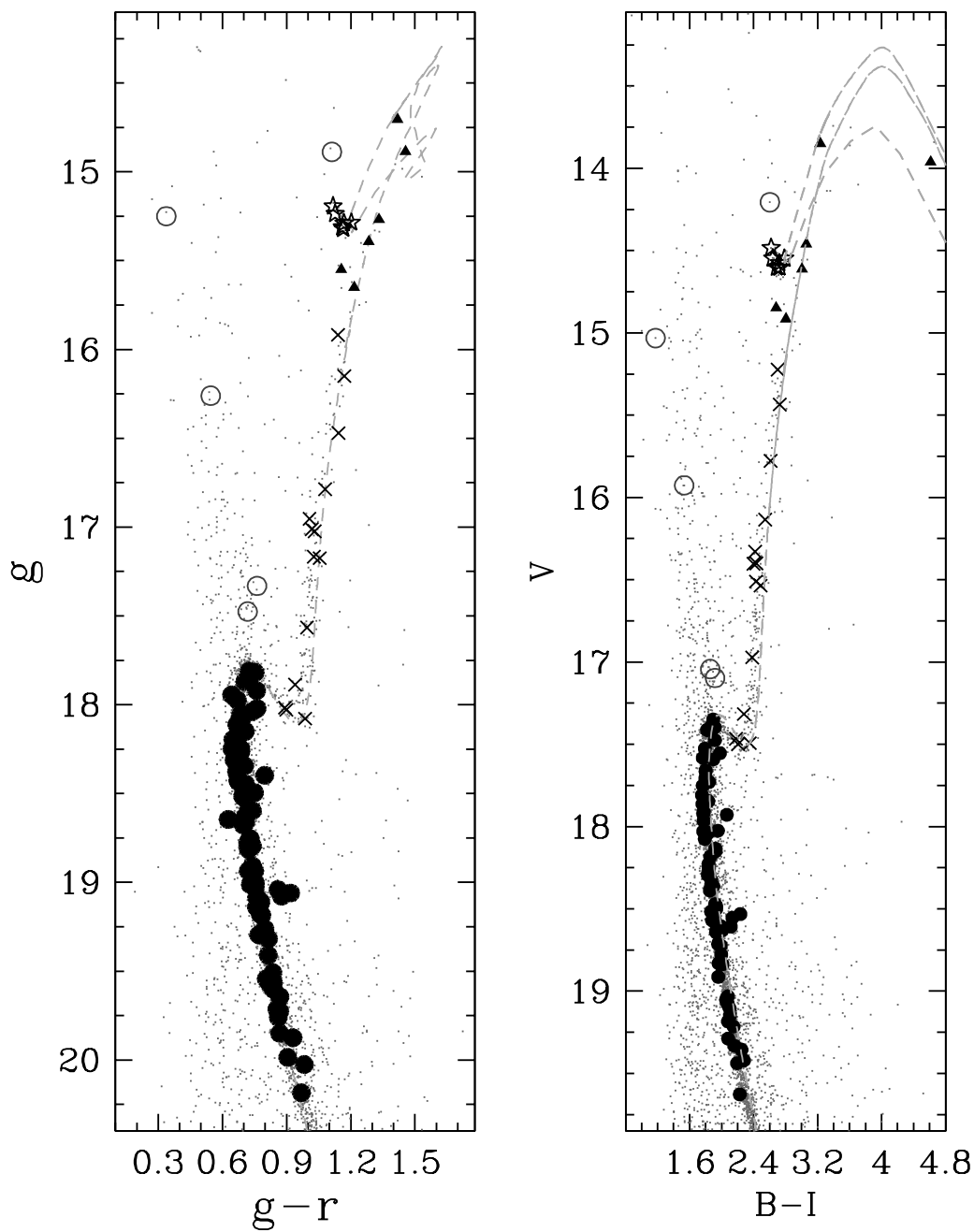}
\caption{Color--magnitude diagrams from \citet[][left]{an2008} and
\citet[][right]{stetson2003}.  The dashed line is an 8
Gyr old isochrone from the BaSTI library (see text for details). Stars at different
evolutionary stages have been plotted with different symbols: MS (filled
circles); lRGB (crosses), RC (open stars), and uRGB (filled triangles). Open
circles are stars selected as NGC~ 6791 members but not included in any of the
evolutionary stages defined.\label{fig_dcm}}
\end{figure}

The SDSS data release also provides information about the magnitudes of the
targets obtained with the SDSS photometric pipeline
\citep{lupton2002,tucker2006,davenport2007}. However, this pipeline was not 
designed to handle the high-density crowded fields typical of the central areas of
clusters. Photometry from SDSS images of stellar clusters, including NGC~6791,
was obtained by \citet{an2008} using the DAOPHOT/ALLFRAME packages 
\citep{stetson1987,stetson1994}. The same packages have been used by
\citet{stetson2003} to derive high-quality broadband $BVI$ 
photometry for this cluster from 1764 CCD images. The resulting color--magnitude
diagrams for NGC~6791 in each photometry system are shown in the left and right
panels of Figure~\ref{fig_dcm}, respectively.  Owing to the higher quality and lower uncertainty of the $BVI$ 
photometry derived by \citet{stetson2003}, we  use these magnitudes  in this paper. The 
position of  stars selected in these diagrams have been used to separate them into
four groups according to their evolutionary stages: 71 MS (filled
circles); 14 lower RGB 
(lRGB, crosses); 6 upper RGB (uRGB, filled triangles); and 6 RC (open stars).
Five stars were discarded because they do not fall within any of these four
groups (open circles). The position of the RGB bump in an 8 Gyr old 
and $Z\sim 0.04$ isochrone (gray dashed line), selected from the BaSTI library
\citep{pietrinferni2004}, was used to separate stars in the lRGB and uRGB.
We divided RGB stars in these two subgroups, because 
no pollution from material synthesized in the stellar interior is expected
before the stars reach the bump. Coordinates, identification numbers, and magnitudes for selected stars are summarized in Table~\ref{tbl-1}.

\section{Index definition}\label{indexdefinition}

The star-to-star light element abundance variations within globular clusters is
investigated from the variations in the strength of the CN absorption bands
around 3839 and 4142 \AA, respectively, and the CH 
band around 4300 \AA~\citep[e.g.][]{norris1979,harbeck2003,kayser2008,pancino2010b}. The strength of
each molecular band is measured with a spectral index defined as the
magnitude of the difference between the 
integrated flux within a wavelength window containing the given feature and the
integrated flux inside a nearby window, or windows, used to define the continuum. Several
definitions of these indexes, used to measure the strength of each molecular band as a function of the luminosity class of the targets,  can been found in the literature.
This is particularly true for the 3839 CN band, where the presence of other
temperature-dependent absorption lines, such as the H$_\zeta$ at 3839
\AA , 
complicates the definition of the continuum window. Since our targets cover a wide luminosity range, we used two different
indexes to measure the strength of the CN band at 
3839 \AA. The first one was defined by \citet{norris1981} to sample RGB stars as:

\begin{equation}
S3839_N=-2.5\log\frac{F_{3846-3883}}{F_{3883-3916}}
\end{equation}

The second one was defined by \citet{harbeck2003} to sample properly MS objects as:

\begin{equation}
S3839_H=-2.5\log\frac{F_{3861-3884}}{F_{3894-3910}}
\end{equation}

The 4142 CN band becomes useful for sampling CN variations in the case of the
relatively metal-rich ([Fe/H]$\sim$-0.7 dex) globular cluster 47
\citep{norris1979}. \citet{pancino2010b} show that this 
band is less sensitive to  CN variations for metal-poor MS stars. However, since NGC~6791 is much more metal-rich
than globular clusters, we also studied the behavior of the
the strength of this molecular band determined as \citep{norris1979}:

\begin{equation}
S4142=-2.5\log\frac{F_{4120-4216}}{F_{4216-4290}}
\end{equation}

Finally, the G CH band at $\sim$4300 \AA~ is typically used for sampling the carbon
abundances in comparison with the CN bands which are used as indicators of N
abundances. We measure the strength of this band 
using the index defined by \cite{lee1999}:

\begin{equation}
CH4300=-2.5\log\frac{F_{4270-4320}}{0.5 F_{4230-4260}+0.5 F_{4390-4420}}
\end{equation}

The adopted windows for each index have been overplotted in
Figure~\ref{fig_spectra}. For each index, the uncertainty has been calculated as
in \citet{pancino2010b} assuming  pure photon (Poisson) noise statistics in the
flux measurements. The
indexes and errors determined together with photometric magnitudes for each
selected star are listed in Table~\ref{tbl-1}. The median
uncertainties for each index and evolutionary stage are listed in
Table~\ref{tbl-2}.

As  mentioned above, two instrumental configurations were 
employed for the NGC~6791
region. A total of 33 stars were observed in both configurations although neither of them met
the criteria used above to select NGC~6791 members. In any case, they are useful for
checking the homogeneity of our data and provided an additional estimate of the
uncertainties. To do this, each index was measured separately in the two spectra
for each star. The medians of the differences between the values obtained for each
index for all stars are: $\Delta (S3839_N)=0.00\pm0.02$; $\Delta
(S3839_H)=0.01\pm0.02$; $\Delta (S4142)=0.000\pm0.006$; and $\Delta
(CH4300)=-0.006\pm0.005$. These differences are lower than the uncertainties
obtained from the Poisson statistics.

\begin{figure}
\includegraphics[scale=0.8]{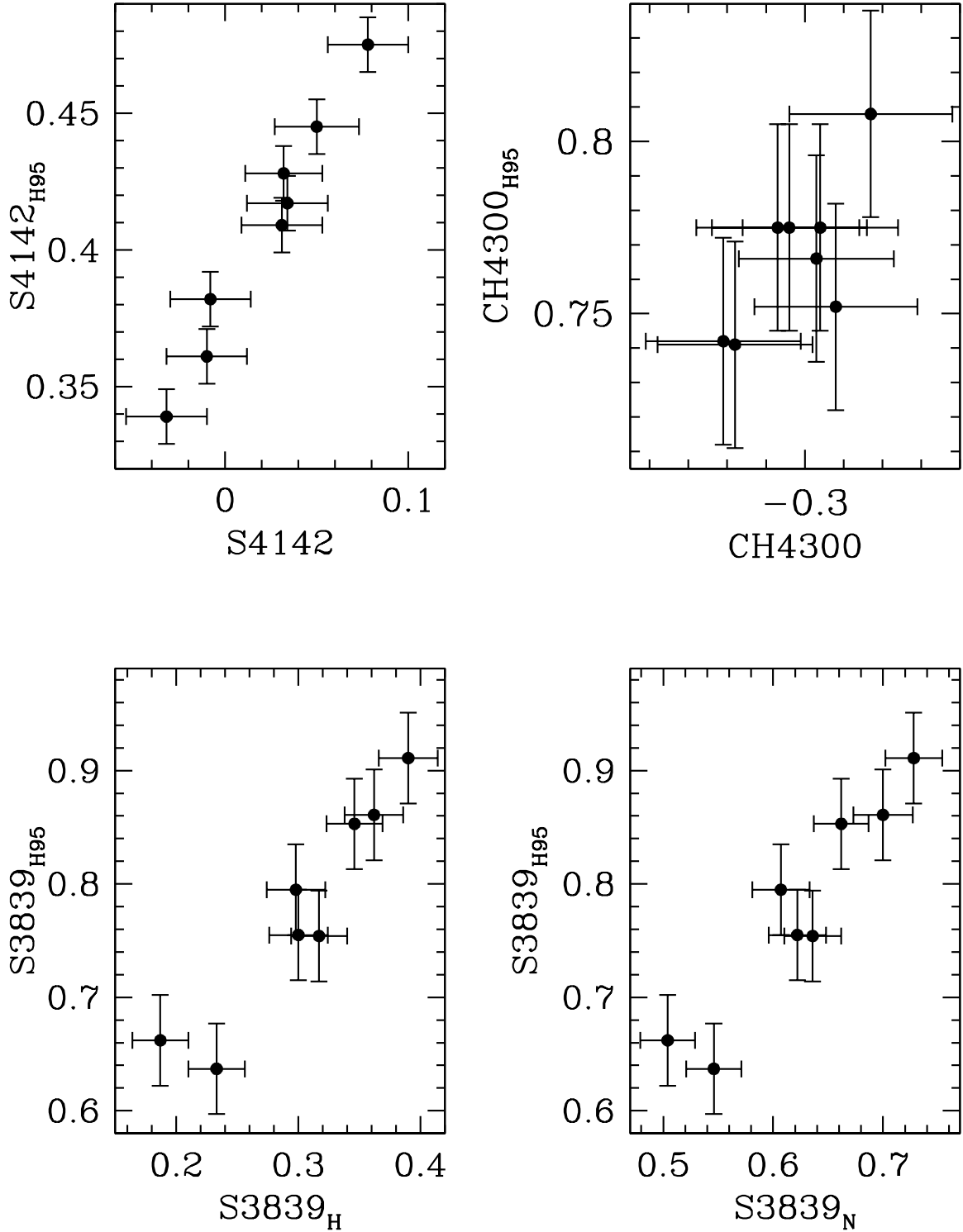}
\caption{Comparison between the indexes measured here with those calculated by \citet{hufnagel1995} for stars in common.\label{fig_comp_hufnagel}}
\end{figure}

\citet{hufnagel1995} studied the CN and CH molecular band strengths in several
RGB and RC stars of NGC~6791 from low resolution spectra (R$<$1000). Our sample has eight
stars in common with that study. Although \cite{hufnagel1995} did not use the
same index definitions used here, a comparison between both sets of
measures  reveals systematic effects due to different instruments used in each
case. The values obtained here against those determined by \citet{hufnagel1995}
for each index have been plotted in Figure~\ref{fig_comp_hufnagel}. A clear linear
correlation is observed for the CN indexes,  particularly in the
case of the S4142 index. This is because we used the same windows to define the
continuum and band intensity as \citet{hufnagel1995}, although they obtained
their indexes in a slightly different way. For each star we have calculated the
difference between the value obtained here and that obtained by
\citet{hufnagel1995}. We have computed the standard deviation of the
differences of all stars for given indexes. The values obtained are: 0.039,
0.036, 0.010, and 0.015 for $S3839_{H}$, $S3839_{N}$, $S4142$ and $CH4300$,
respectively. These values are of the order of the median of the uncertainties
of each index for RGB or RC stars (see Table~\ref{tbl-2}).

\section{CN and CH distributions}\label{CNCHdistributions}

\begin{figure}
\includegraphics[scale=0.7]{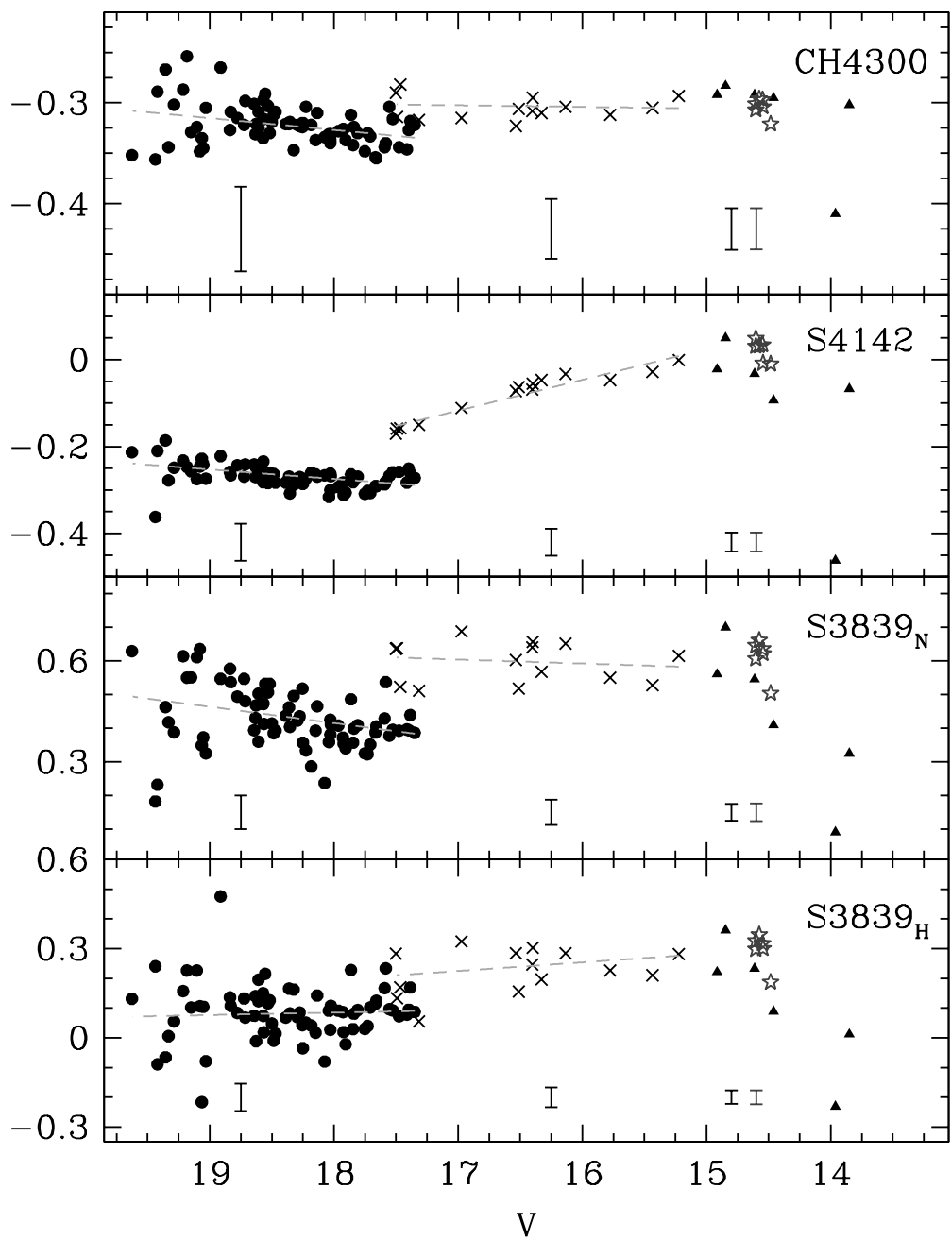}
\caption{Run of the four molecular indexes versus $V$ magnitudes. Stars at
different evolutionary
stages have been plotted with different symbols as in Figure~\ref{fig_dcm}.
Dashed lines are the linear least-squares fit to those stars in the MS and lRGB
separately used to correct the temperature and gravity dependence (see text for
details). Uncertainties obtained as the median of the errors of each index 
determination for each evolutionary stage have been plotted in the bottom part of each panel.\label{fig_index_V}}
\end{figure}

The run of the four molecular indexes defined above as a function of $V$
magnitude
are shown in Figure~\ref{fig_index_V}. For MS stars, the strength decreases as
the $V$
magnitude increases. Since  temperature increases with magnitude for MS
objects, the observed trend is explained by the more efficient formation of CN
and CH molecules at lower temperatures. For lRGB objects, the behaviour is more
random: as we ascend the RGB towards brighter magnitudes for $V\gtrsim$15 the
indexes $S3839_{H}$ and $S4142$ increases,  the index $S3839_{N}$ slightly
decreases, and $CH4300$ remains almost constant.
From $V\sim$15 a possible turnover is observed in the strengths of the indexes 
studied. We will come back to this point in Section \ref{discussion}. 

\subsection{Main Sequence and lower Red Giant Branch}

First, we are going to focus on those stars in the MS and lRGB. As was
explained above, according to  stellar evolution models, it is expected that
the chemical
composition of their atmospheres may reflect the initial conditions of the
molecular gas cloud from which the stars were formed. Before using CN and CH
bands as chemical composition indicators, the temperature and gravity dependence
of molecular band strengths
should be removed. Different approaches can be found in the literature to removing
this dependence, such as fitting the lower envelope or the median
ridge line of a given molecular index as a function of color or magnitude.
However, due to the low number of objects sampled in the case of the lRGB,
this step should be performed with caution. To surmount this problem we adopted the
following procedure. A linear least-squares fit was performed on half of the
stars (randomly selected) in the lRGB. This procedure was repeated
10$^3$ times using different random subsets each time. The final zero-point and
slope for each index were obtained as the median of the zero points and slopes calculated in
each individual test. The same procedure has been used for MS stars. With this
procedure we tried to minimize the uncertainties due to the low number of
object studied and the influence of the points on the edges. The final linear fit adopted in each case is shown as dashed
lines in Figure~\ref{fig_index_V} and are listed in
Table~\ref{tbl-3}. We calculated the corrected pseudo-indexes, denoted by a  $\delta$
preceding the corresponding index, as the difference
between the index and the adopted linear fit in each case.

\begin{figure}
\includegraphics[scale=0.7]{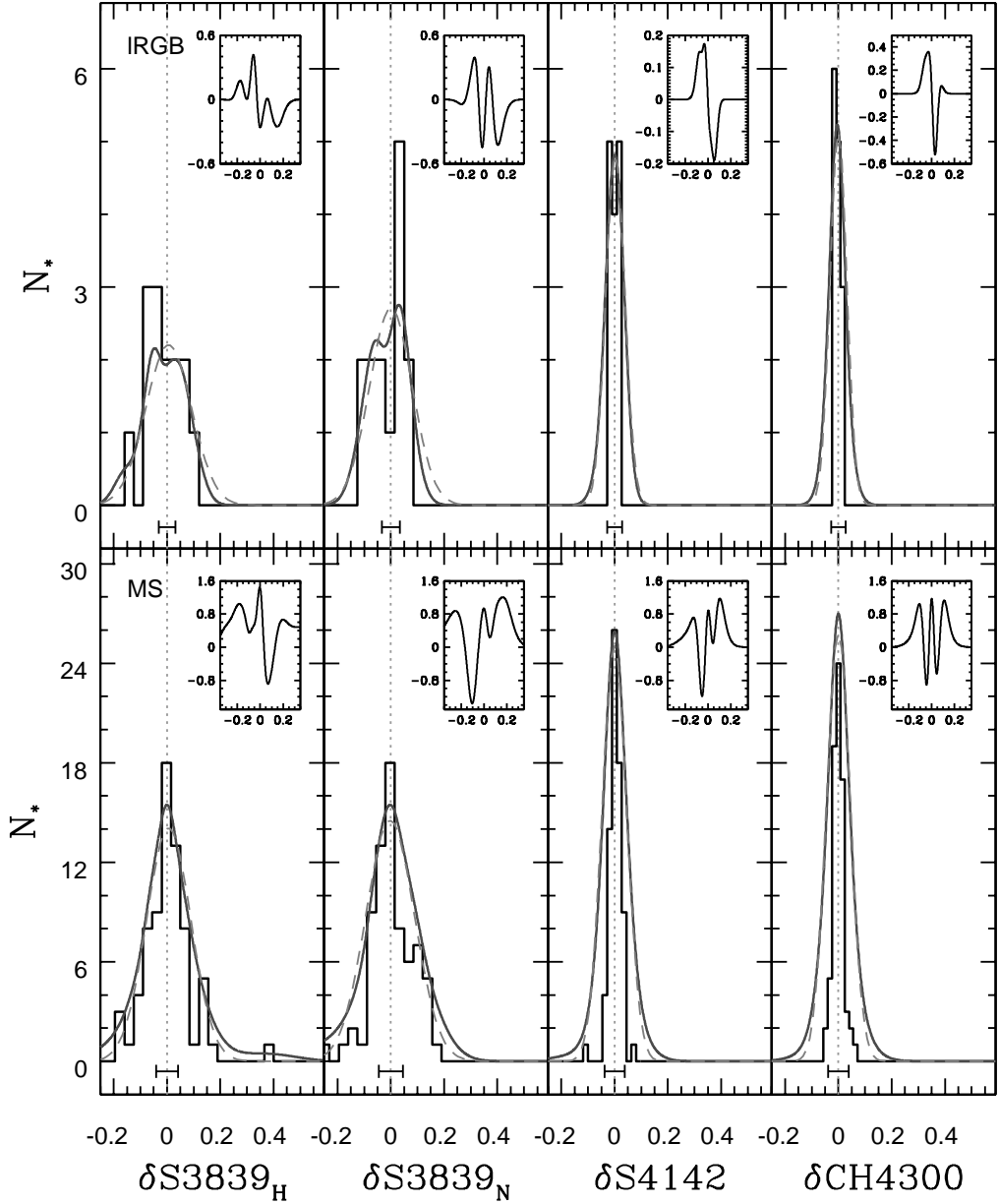}
\caption{Histograms and generalized histograms (solid lines) for each index
obtained for lRGB (top) and MS (bottom) stars, respectively. As a comparison, each
generalized distribution has been fitted with a single Gaussian (dashed lines)
and the residual between them are shown in inset panels. The residuals suggest
that a single Gaussian does not  properly reproduce the observed distributions.\label{fig_histo}}
\end{figure}

The  histograms obtained for each corrected index are shown in Figure~\ref{fig_histo} for the MS (bottom) and lRGB (upper), respectively. In each case, generalized
histograms (solid lines) have been derived by assuming that each
star is represented by a Gaussian probability function centered on its corrected
index value whose $\sigma$ is equal to the uncertainty in the
determination of the index. There is no hint of bimodalities for any of the
four indexes in the case of the MS. However, the distribution widths of the two
S3839 indexes seem larger than that expected from the median uncertainties, as shown in the bottom of each panel, 
which in both cases is about 0.04 (Table~\ref{tbl-2}). In fact, if we fitted each generalized
histogram with a single Gaussian (dashed lines) the
$\sigma$ values obtained are 0.083$\pm$0.001 and 0.091$\pm$0.001 for $S3839_{H}$
and $S3839_{N}$, respectively. These values are twice as great as the
uncertainties. In contrast, narrow distributions are obtained for $S4142$ and
$CH4300$. Although the median uncertainties are similar to those of
S3839 indexes ($\sim$0.04). In this case, the $\sigma$ obtained from the fit of
a single Gaussian are 0.046$\pm$0.001 and  0.045$\pm$0.001, respectively. The
residuals between the generalized histogram (solid lines) and the
single Gaussian fitted (dashed lines) have been plotted in inset panels. They
suggest that the distributions are not well reproduced by a single Gaussian.

For lRGB objects, the S4142 and CH4300 indexes, with a median uncertainty of
$\sim$0.03, also show a narrow distribution. They are reasonably reproduced by
single Gaussians (dashed lines) with $\sigma\sim$0.037$\pm$0.001 and
0.031$\pm$0.001 for $S4142$ and $CH4300$, respectively. However, the residuals
suggest that the generalized histograms have a wider dispersion than that
obtained from the single Gaussian fit. In the case of
two S3839 indexes a bimodality is sensed both in the normal and generalized
histograms. Again, we have tried to fit each generalized distribution with a
single Gaussian. In this case, the $\sigma$ values obtained are 0.086$\pm$0.001 and
0.080$\pm$0.001, for $S3839_{H}$
and $S3839_{N}$, respectively. These values are almost three times the median
uncertainties ($\sim$0.03) in each case. Again, the residuals of the difference
between the generalized histogram and the single Gaussian fit suggest that a single
Gaussian does not properly
reproduce  the observed distributions.

\begin{figure}
\includegraphics[scale=0.7]{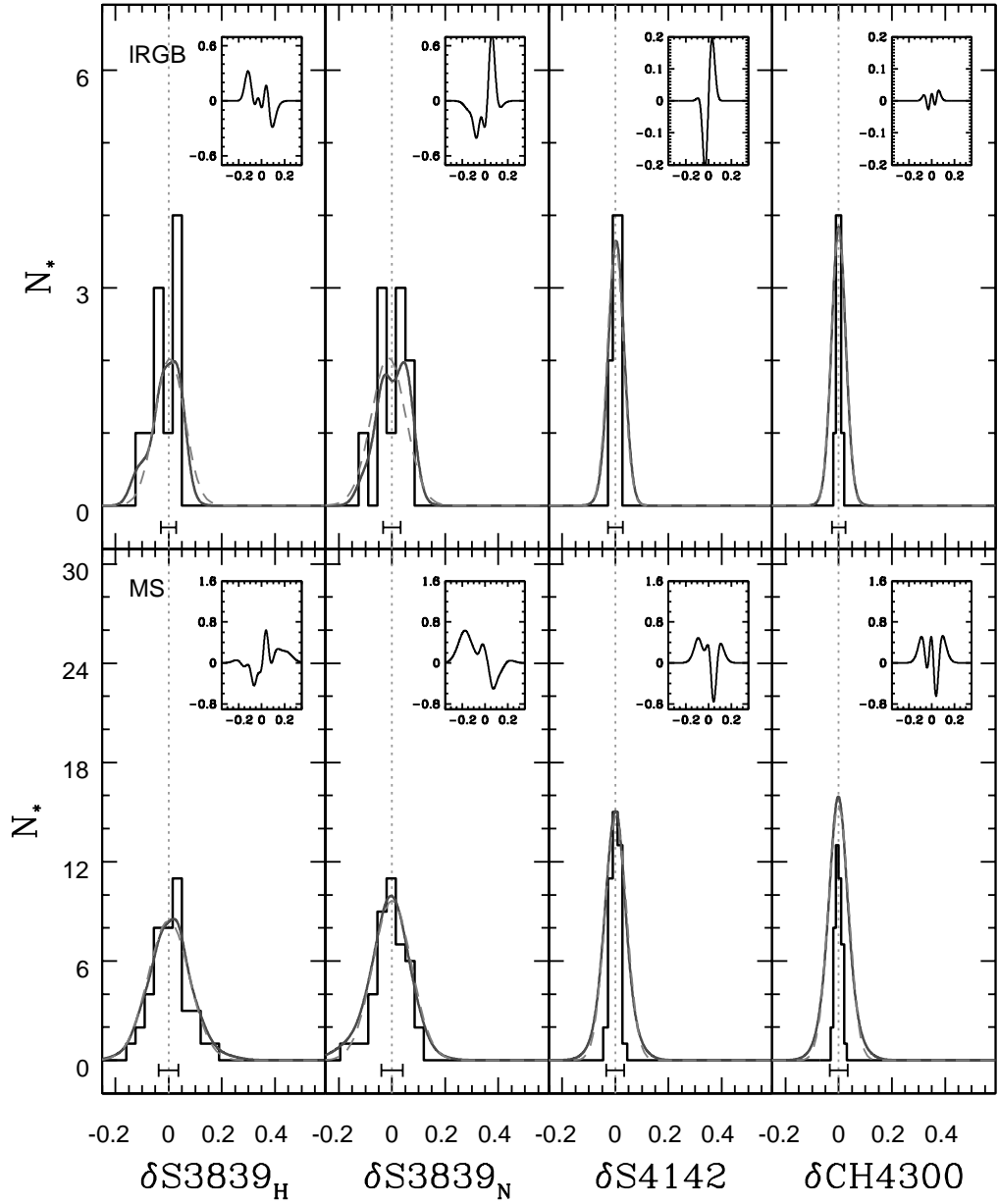}
\caption{As Figure~\ref{fig_histo} but for lRGB stars with V$\leq$17.25 and MS objects in the range 18.75$\leq$V$\leq$17.75.\label{fig_histo_test}}
\end{figure}

It can be argued that the wider distributions observed in the case of
the $S3839$ indexes are due to our assumption of a linear dependence of the
molecular band strength with temperature and gravity. This approximation may be
wrong in the case of the MS turn-off and the base of the RGB. For this reason
we have repeated our analysis eliminating lRGB stars fainter than V$\sim$17.25
mag and MS objects brighter than V$\sim$17.75 mag. Moreover, we have rejected
MS stars fainter than V$\sim$18.75 mag owing to the larger uncertainties in the
index determinations. The distributions of each index for the stars at each evolutionary
stage are shown in Figure~\ref{fig_histo_test}. They have been obtained following the same 
procedure as described previously but using the restricted samples to perform the least-squares 
fits in order to remove the temperature and gravity dependence. In this case, the signs of 
bimodalities observed in the histograms of $S3839$ indexes for lRGB stars seem  clearer. However, these signs are smoothed when the generalized histograms
are obtained taking into account the uncertainties. As before, the distributions
are wider than those expected from the uncertainties. However, in this case the same
behavior is also observed in the case of $S4142$ and $CH4300$ indexes. We
have fitted each generalized histogram with a single Gaussian (dashed lines) and
plotted the residuals in the inset panels. Again, the residuals suggest that the
distributions are not well reproduced by a single Gaussian with the exception of the $CH4300$ index. Therefore, we
conclude that the behaviors described above for each index are not influenced
by the assumption of a linear dependence of the molecular index strengths with
temperature and gravity for stars near the MS turn-off and in the base of the
RGB. Moreover, it seems that the strengths of the $S4142$ band have the same behavior. On contrary, the CH band shows no spread.

\subsection{Red Clump}

\begin{figure}
\includegraphics[scale=0.7]{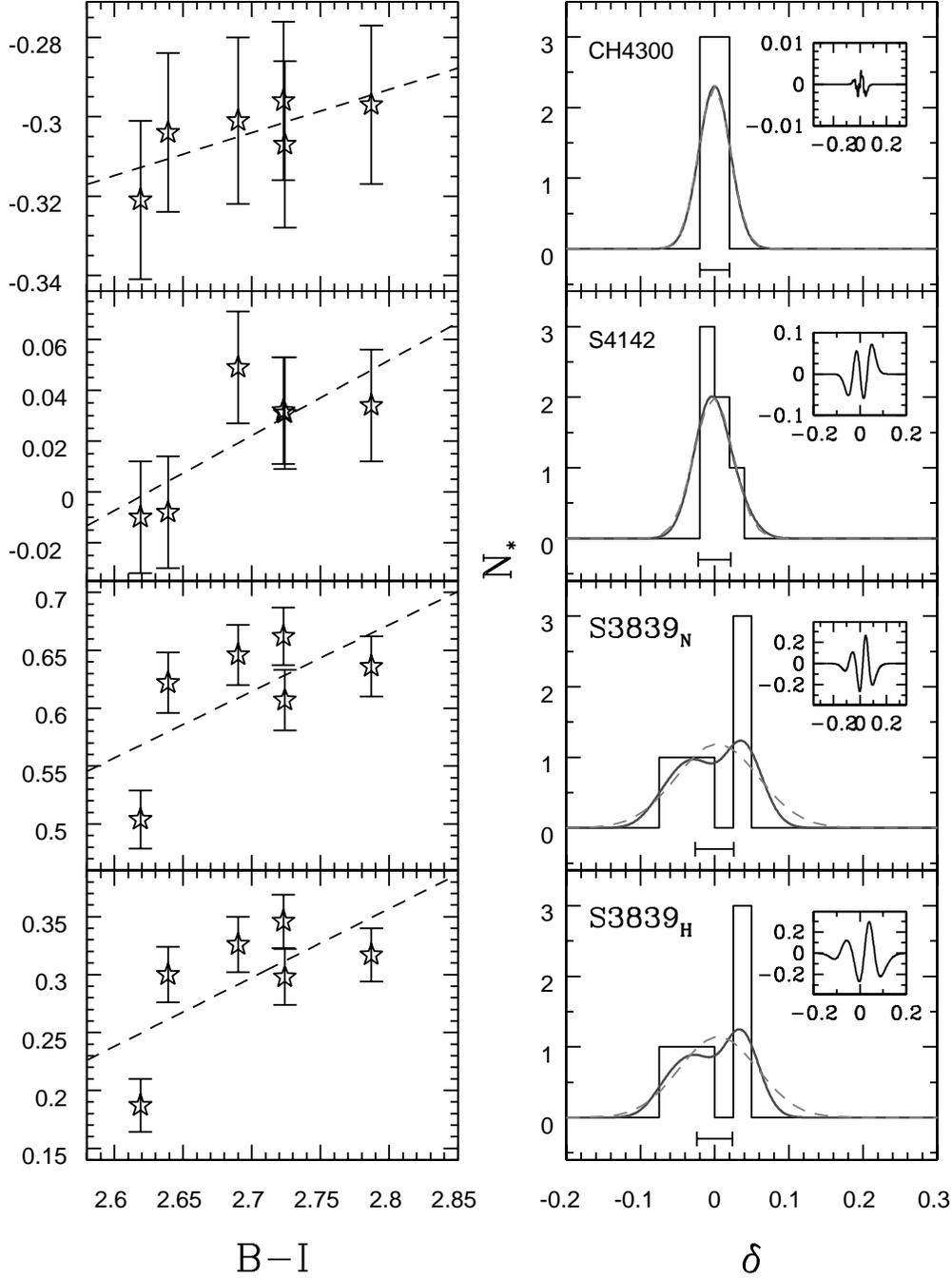}
\caption{Left: run of the strength of the molecular indexes of RC stars versus
their $B-I$ colors. A line has been fitted (dashed line) in order to remove the
temperature and gravity dependence in order to obtain the corrected 
pseudo-indexes. Right: Normal (histogram) and generalized (solid line) distributions of
each corrected pseudo-index. Dashed lines are the best single Gaussians fitted to
each generalized distribution. The residuals between the single Gaussian and the
generalized distribution are shown in inset panels.\label{fig_RC}}
\end{figure}

As discussed above, \citet{hufnagel1995} studied 31 RC stars in a similar way to that described here. They found inhomogeneties of the CN bands but no
signs of bimodalities as observed among globular clusters. In our sample, we
have six stars in the RC region, five of them having also been studied by
\citet{hufnagel1995}. In the case of RC stars, the $V$ magnitude is not useful
for  removing the temperature and gravity dependence since they all have similar
values. For this reason, in left panels of Figure~\ref{fig_RC} each index has been
plotted against the  $B-I$ color. A linear fit has been performed for each index
in order to remove the temperature dependence. The very small number of stars
observed in the RC do not allow us to follow the same procedure used in the case
of MS and lRGB stars. In each case, the corrected index has been obtained as the
difference between the index and the value of the linear fit. The histograms of
the distributions obtained for each corrected index have been plotted in the right
panels of Figure~\ref{fig_RC}. The two S3839 indexes seem to have a bimodal
behavior. However, this bimodality is smoothed when the generalized histogram is
obtained taking into account the uncertainties (the solid lines in the right panels of
Fig.~\ref{fig_RC}). In any case, the generalized histograms seem to be wider
 than that expected from the uncertainties. To investigate this point, we
fitted a single Gaussian to each generalized histogram (dashed lines). The $\sigma$ values
of the single Gaussian fitted are 0.053$\pm$0.005 and 0.056$\pm$0.005 for
$S3839_{H}$
and $S3839_{N}$, respectively. Therefore, the  distributions obtained are twice as
wide as is expected from uncertainties. A similar result was obtained by
\citet{hufnagel1995}.

In the case of the S4142 and CH4300 indexes, the  distributions obtained are well
reproduced by a single Gaussian. Although the S4142 distributions seems slightly
wider ($\sim$0.03) than that  expected from the uncertainties. On the contrary, the CH4300
distribution has a width ($\sim$0.02) similar to the uncertainty. The lack of
variations in the CH band strength for RC stars in NGC~6791 was also reported by
\citet{hufnagel1995}.

\section{Discussion}\label{discussion}

Significant variations and bimodalities in the strengths of the CN molecular
bands, which are always anticorrelated with the CH strengths, have been
widely reported in globular cluster stars at different evolutionary stages
\citep[e.g.][]{cannon1998,cohen1999,harbeck2003,kayser2008,pancino2010b,smolinski2011}. This result is
explained by the existence of  different star-to-star C and N abundances. The recent discovery of several evolutionary sequences in almost
all globular clusters studied \citep[e.g.][]{piotto2007} has associated these chemical
composition inhomogeneties with the existence of several stellar populations in each system.

In contrast, studies on open clusters, which are more metal-rich, younger, and less
massive than the globulars, have not been observed  similar trends
\citep[e.g.][]{norris1985,hufnagel1995,martell2009}. This is interpreted in terms of open
clusters being  formed by a single stellar population. However, similar abundance
variations to those seen among globular stars would produce smaller scatter in
the CN band strengths of open cluster stars owing to their higher metallicities, with solar
or above solar metal content \citep{hufnagel1995}. Therefore, the same trends
observed in globulars would be more difficult to detect in open clusters.
  
In this sense, NGC~6791 is a key system because it has a mass intermediate between globular
and open clusters. \citet{hufnagel1995} reported an unexpected
scatter among the strengths of the CN bands in the RC stars of NGC~6791. We confirm
this spread not only in evolved RC stars but
also among MS and lRGB objects. Therefore, the different abundances responsible
for this spread
 should be present in the molecular gas cloud from which NGC~6791 stars
were formed. For example, the scatter observed in the strengths of CN band at
3839 \AA~is twice that expected from the uncertainties.
Moreover, some signs of bimodalities are sensed in the distributions of the
indexes used to
measure the strength of this band among lRGB and RC stars. However, this result
should be treated with caution owing to the the small number of stars studied in each
case. We do not detect the CH--CN band strength anticorrelations observed in
globular clusters, and all the stars studied seem to
have a very similar CH band strength. 


It is difficult to explain the CN strength dispersion
observed in terms of mixing effects since this trend is also observed among unevolved
stars. Moreover, \citet{geisler2012} have recently reported an intrinsic dispersion
of Na and O abundances among NGC~6791 stars in the whole RGB. These stars follow
the same Na--O anticorrelation observed in globular clusters. For analogy with
globular clusters, these results can be explained by the existence of multiple
stellar populations,  NGC~6791 being the first open cluster to show this feature. In
fact,the unexpectedly
wide RGB and MS observed in the NGC~6791 color--magnitude diagram indicates an
age spread of about 1 Gyr \citep{twarog2011}. However, these features could also be
 due to the existence of
differential reddening \citep{platais2011}.

Recent investigations have demonstrated that light element abundances are
correlated with the colors of RGB stars if near ultraviolet filters (i.e.\ 
Johnson $U$ or SDSS $u$) are used  in globular clusters \citep[e.g.][]{marino2008,milone2010,milone2012,lardo2012}, although the details of this
behavior are still unclear. An accurate
wide-field photometric study including $U$ filter would therefore provide information about the radial distribution of the different
stellar populations, which are key to investigating the process of the cluster 
formation and early chemical enrichment \citep[e.g.][]{decresin2007,dercole2008,renzini2008}.

Finally, the decrease in the strengths of the CN bands with magnitude 
for $V \lesssim 15$  needs  further consideration although it is beyond the scope of this paper. A similar trend
has been observed in some more metal-poor globular clusters by
\citet{smolinski2011}. This turnover may not be unusual since it has been
observed by other tracers of  CN band strengths, such as DDO photometry,
in other open cluster; e.g.\ NGC~188 \citep{mcclure1974}, NGC~2682
\citep{janessmith1984}, and NGC~7789 \citep{janes1977}. However, this question
has not been studied in detail by any of these authors. A detailed study of
DDO colors in stars of  relatively metal-rich globular
clusters shows that this decrease in  CN strength is observed only for
CN-rich stars and  not for  CN-weak objects.
They compared their observations with colors obtained from synthetic spectra
and concluded that, although it is expected that the mixing with material
synthesized in the interior by  the CN cycle should 
increase the N abundances, and therefore the CN strengths, the depletion of C
would be so large that it would produce the decrease in the CN strengths
observed, even though C is the lesser contributor to the formation of the CN
molecule. In any case, a detailed analysis of this issue is necessary in order to understand the process that produces it.

\section{Summary}\label{summary}

We have studied the strengths of the CN and CH bands at
3839, 4142, and 4300 \AA~ in stars at different
evolutionary stages in the open cluster NGC~6791. This system is one of the most
massive open clusters known, and according to its orbit, it should have been more
massive at the time of its formation. For this analysis, we used low-resolution
spectra ($R \sim 2000$) obtained in the framework of the SEGUE project within the
SDSS. Our main results are as
follows:

\begin{itemize}
\item A significant spread in the strengths of the CN molecular band at 3839
\AA~ is observed not only in evolved red clump stars but also among objects in
the main sequence and lower red giant branch. This dispersion is at least twice
as great as   that expected from the uncertainties. This result is obtained by
the two indexes used to measure the strength of this band. This scatter is not
observed in the strengths of the CN band at 4142 \AA. 
\item The distributions of the strengths of the two CN bands studied are not
well reproduced by a single Gaussian even for main sequence stars. In fact, signs
of bimodalities appear when the obtained distribution is compared with a single
Gaussian.
\item No significant dispersion is observed in the strengths of the CH band at
4300 \AA. In fact, the distributions of the strengths of these indexes are
relatively well reproduced by a single Gaussian. This is particularly true in
the case of red clump stars.  
\item The strengths of the three molecular bands studied decreases as the
magnitude of the star is brighter in the upper part of the red giant branch
above the red clump.
\end{itemize}

\acknowledgments

I am indebted to the anonymous referee for his/her comments and suggestions
which have improved significantly the analysis and the results presented in this
paper. Comments on an earlier version of the manuscript by A. Aparicio, C.
Gallart, M. Monelli, and N. Ospina are warmly acknowledge. R. C. acknowledges 
funds provided by the Spanish
Ministry of Science and Innovation under the Juan de la Cierva fellowship and
under the Plan Nacional de Investigaci\'on Cient\'{\i}fica,
Desarrollo, e Investigaci\'on Tecnol\'{\i}gica, AYA2010-16717. This research has
made use of the WEBDA database, operated at the Institute for Astronomy of the
University of Vienna, and the SIMBAD database,
operated at the CDS, Strasbourg, France

Funding for SDSS-III has been provided by the Alfred P. Sloan Foundation, the
Participating Institutions, the National Science Foundation, and the U.S.
Department of Energy Office of Science. The SDSS-III web site is
http://www.sdss3.org/. SDSS-III is managed by the Astrophysical Research
Consortium for the Participating Institutions of the SDSS-III Collaboration
including the University of Arizona, the Brazilian Participation Group,
Brookhaven National Laboratory, University of Cambridge, University of Florida,
the French Participation Group, the German Participation Group, the Instituto de
Astrofisica de Canarias, the Michigan State/Notre Dame/JINA Participation Group,
Johns Hopkins University, Lawrence Berkeley National Laboratory, Max Planck
Institute for Astrophysics, New Mexico State University, New York University,
Ohio State University, Pennsylvania State University, University of Portsmouth,
Princeton University, the Spanish Participation Group, University of Tokyo,
University of Utah, Vanderbilt University, University of Virginia, University of
Washington, and Yale University.

{\it Facilities:} \facility{SDSS(SEGUE)}.




\clearpage

\begin{deluxetable}{cccccccccccccl}
\rotate
\tabletypesize{\scriptsize}
\tablecaption{Index measurements for the sample stars.\label{tbl-1}}
\tablewidth{0pt}
\tablehead{
\colhead{Plate} & \colhead{Fiber} & \colhead{RA} & \colhead{Dec} & \colhead{g} & \colhead{g-r} & \colhead{ID$_{Stetson}$} & \colhead{V} &\colhead{V-I} & \colhead{$S3839_H$} & \colhead{$S3839_N$} & 
\colhead{$S4142$} & \colhead{$CH4300$} & \colhead{Region}
}
\startdata
2800  &  150  &  19:21:04.28  & +37:47:18.84  & 14.70  &  1.42  &  11814  & 13.85  &  1.66  &  0.01$\pm$0.02  &  0.33$\pm$0.02  & -0.07$\pm$0.02  & -0.30$\pm$0.02  & uRGB\\
2800  &  151  &  19:21:14.52  & +37:46:32.81  & 17.89  &  0.94  &  14109  & 17.32  &  1.14  &  0.05$\pm$0.04  &  0.51$\pm$0.05  & -0.15$\pm$0.04  & -0.32$\pm$0.04  & lRGB\\
2800  &  152  &  19:21:06.70  & +37:48:08.31  & 18.63  &  0.71  &  12451  & 18.18  &  0.94  &  0.04$\pm$0.07  &  0.29$\pm$0.07  & -0.26$\pm$0.06  & -0.32$\pm$0.06  & MS\\
2800  &  154  &  19:21:01.45  & +37:48:05.11  & 16.47  &  1.14  &  11006  & 15.78  &  1.31  &  0.23$\pm$0.03  &  0.55$\pm$0.03  & -0.05$\pm$0.02  & -0.31$\pm$0.02  & lRGB\\
\enddata
\tablecomments{Table \ref{tbl-1} is published in its entirety in the 
electronic edition of the {\it Astrophysical Journal}.  A portion is 
shown here for guidance regarding its form and content.}
\end{deluxetable}

\begin{deluxetable}{lccccc}
\tablecaption{Median uncertainties for each index and evolutionary stage.\label{tbl-2}}
\tablewidth{0pt}
\tablehead{
\colhead{Region} & \colhead{$S3839_H$} & \colhead{$S3839_N$} & \colhead{$S4142$} & \colhead{$CH4300$} 
}
\startdata
MS   & 0.041$\pm$0.002 &  0.045$\pm$0.002 & 0.038$\pm$0.002 & 0.038$\pm$0.002\\
lRGB & 0.031$\pm$0.002 &  0.034$\pm$0.003 & 0.028$\pm$0.002 & 0.027$\pm$0.002\\
uRGB & 0.023$\pm$0.005 &  0.025$\pm$0.001 & 0.022$\pm$0.003 & 0.021$\pm$0.002\\
RC   & 0.024$\pm$0.001 &  0.026$\pm$0.001 & 0.022$\pm$0.001 & 0.020$\pm$0.001\\
\enddata
\end{deluxetable}

\begin{table}
\begin{center}
\caption{Coefficients of the linear fit used to construct the corrected pseudo-indexes 
and marked as dashed lines in Figs~\ref{fig_index_V} and \ref{fig_RC}, respectively.\label{tbl-3}}
\begin{tabular}{lcccccc}
\tableline\tableline
Index & \multicolumn{2}{c}{Main-Sequence\tablenotemark{a}} & \multicolumn{2}{c}{lower Red Giant Branch\tablenotemark{a}} & \multicolumn{2}{c}{Red Clump\tablenotemark{b}}  \\
 & zp & slope & zp & slope & zp & slope  \\
\tableline
S3839$_H$ & 0.23$\pm$0.06 & -0.008$\pm$0.003 & 0.67$\pm$0.20 & -0.03$\pm$0.01 & -1.39$\pm$0.25 & 0.59$\pm$0.03 \\
S3839$_N$ & -0.46$\pm$0.07 & 0.049$\pm$0.004 & 0.42$\pm$0.17 & 0.01$\pm$0.01 & -0.94$\pm$0.26 & 0.57$\pm$0.04 \\
S4142          & -0.64$\pm$0.02 & -0.020$\pm$0.001 & 1.08$\pm$0.06 & -0.070$\pm$0.004 & -0.78$\pm$0.03 & 0.296$\pm$0.005 \\
CH4300       & -0.53$\pm$0.01 & -0.011$\pm$0.001 & -0.32$\pm$0.04 & 0.001$\pm$0.001 & -0.597$\pm$0.005 & 0.108$\pm$0.001 \\
\tableline
\end{tabular}
\tablenotetext{a}{Obtained as $index=zp+slope\times V$.}
\tablenotetext{b}{Obtained as $index=zp+slope\times (B-I)$.}
\end{center}
\end{table}


\begin{thebibliography}

\bibitem[Aihara et al.(2011)]{aihara2011} Aihara, H., Allende-Prieto, C., An, D., et al.\ 2011, \apjs, 193, 29 

\bibitem[Allende Prieto et al.(2008)]{allendeprieto2008} Allende Prieto, C., Sivarani, T., Beers, T.~C., et al.\ 2008, \aj, 136, 2070 

\bibitem[An et al.(2008)]{an2008} An, D., Johnson, J.~A., Clem, J.~L., et al.\ 2008, \apjs, 179, 326 


\bibitem[Bedin et al.(2006)]{bedin2006} Bedin, L.~R., Piotto, G., Carraro, G., King, I.~R., \& Anderson, J.\ 2006, \aap, 460, L27 

\bibitem[Bedin et al.(2008)]{bedin2008} Bedin, L.~R., Salaris, M., Piotto, G., et al.\ 2008, \apjl, 679, L29


\bibitem[Briley et al.(2001)]{briley2001} Briley, M.~M., Smith, 
G.~H., \& Claver, C.~F.\ 2001, \aj, 122, 2561 

\bibitem[Buzzoni et al.(2012)]{buzzoni2012} Buzzoni, A., Bertone, E., Carraro, G., \& Buson, L.\ 2012, \apj, 749, 35 

\bibitem[Cannon et al.(1998)]{cannon1998} Cannon, R.~D., Croke, B.~F.~W., Bell, R.~A., Hesser, J.~E., 
\& Stathakis, R.~A.\ 1998, \mnras, 298, 601 


\bibitem[Carrera et al.(2007)]{carrera2007} Carrera, R., Gallart, 
C., Pancino, E., \& Zinn, R.\ 2007, \aj, 134, 1298 

\bibitem[Carrera \& Pancino(2011)]{carrera2011} Carrera, R., \& Pancino, E.\ 2011, \aap, 535, 30

\bibitem[Carraro et al.(2006)]{carraro2006} Carraro, G., Villanova, S., Demarque, P., et al.\ 2006, \apj, 643, 1151 

\bibitem[Carretta et al.(2007)]{carretta2007} Carretta, E., Bragaglia, A., \& Gratton, R.~G.\ 2007, \aap, 473, 129 

\bibitem[Carretta et al.(2009a)]{carretta2009a} Carretta, E., Bragaglia, A., Gratton, R.~G., et al.\ 2009a, \aap, 505, 117 

\bibitem[Carretta et al.(2009b)]{carretta2009b} Carretta, E., Bragaglia, A., Gratton, R., \& Lucatello, S.\ 2009b, \aap, 505, 139 

\bibitem[Carretta et al.(2010)]{carretta2010} Carretta, E., Bragaglia, A., Gratton, R.~G., et al.\ 2010, \aap, 516, A55 

\bibitem[Cohen(1999)]{cohen1999} Cohen, J.~G.\ 1999, \aj, 117, 2434 



\bibitem[Da Costa \& Cottrell(1980)]{dacosta1980} Da Costa, G.~S., \& Cottrell, P.~L.\ 1980, \apjl, 236, L83 

\bibitem[Davenport et al.(2007)]{davenport2007} Davenport, J.~R.~A., 
Bochanski, J.~J., Covey, K.~R., et al.\ 2007, \aj, 134, 2430 


\bibitem[D'Ercole et al.(2008)]{dercole2008} D'Ercole, A., Vesperini, E., D'Antona, F., McMillan, S.~L.~W., \& Recchi, S.\ 2008, \mnras, 391, 825 

\bibitem[de Silva et al.(2009)]{desilva2009} de Silva, G.~M., Gibson, B.~K., Lattanzio, J., \& Asplund, M.\ 2009, \aap, 500, L25

\bibitem[Decressin et al.(2007)]{decresin2007} Decressin, T., Meynet, G., Charbonnel, C., Prantzos, N., \& Ekstr{\"o}m, S.\ 2007, \aap, 464, 1029 

\bibitem[Geisler et al.(2012)]{geisler2012} Geisler, D., Villanova, S., Carraro, G., et al.\ 2012, arXiv:1207.3328 

\bibitem[Gratton et al.(2006)]{gratton2006} Gratton, R., Bragaglia, 
A., Carretta, E., \& Tosi, M.\ 2006, \apj, 642, 462 

\bibitem[Gratton et al.(2010)]{gratton2010} Gratton, R.~G., Carretta, E., Bragaglia, A., Lucatello, S., \& D'Orazi, V.\ 2010, \aap, 517, A81 


\bibitem[Harbeck et al.(2003)]{harbeck2003} Harbeck, D., Smith, G.~H., \& Grebel, E.~K.\ 2003, \aj, 125, 197 

\bibitem[Hufnagel et al.(1995)]{hufnagel1995} Hufnagel, B., Smith,  G.~H., \& Janes, K.~A.\ 1995, \aj, 110, 693 

\bibitem[Janes(1984)]{janes1984} Janes, K.~A.\ 1984, \pasp, 96, 977

\bibitem[Janes(1977)]{janes1977} Janes, K.~A.\ 1977, \aj, 82, 35 

\bibitem[Janes \& Smith(1984)]{janessmith1984} Janes, K.~A., \& Smith, G.~H.\ 1984, \aj, 89, 487 

\bibitem[Kalirai et al.(2007)]{kalirai2007} Kalirai, J.~S., Bergeron, P., Hansen, B.~M.~S., et al.\ 2007, \apj, 671, 748 

\bibitem[Kayser et al.(2008)]{kayser2008} Kayser, A., Hilker, M., Grebel, E.~K., \& Willemsen, P.~G.\ 2008, \aap, 486, 437

 
\bibitem[Lardo et al.(2012)]{lardo2012} Lardo, C., Milone, A.~P., Marino, A.~F., et al.\ 2012, \aap, 541, A141 

\bibitem[Lee(1999)]{lee1999} Lee, S.-G.\ 1999, \aj, 118, 920 

\bibitem[Lee et al.(2008a)]{lee2008a} Lee, Y.~S., Beers, T.~C., Sivarani, T., et al.\ 2008a, \aj, 136, 2022 

\bibitem[Lee et al.(2008b)]{lee2008b} Lee, Y.~S., Beers, T.~C., Sivarani, T., et al.\ 2008b, \aj, 136, 2050 

\bibitem[Lupton et al.(2002)]{lupton2002} Lupton, R.~H., Ivezic, 
Z., Gunn, J.~E., et al.\ 2002, \procspie, 4836, 350 

\bibitem[Marino et al.(2008)]{marino2008} Marino, A.~F., Villanova, S., Piotto, G., et al.\ 2008, \aap, 490, 625 

\bibitem[Martell \& Smith(2009)]{martell2009} Martell, S.~L., \& Smith, G.~H.\ 2009, \pasp, 121, 577 

\bibitem[McClure(1974)]{mcclure1974} McClure, R.~D.\ 1974, \apj, 194, 355 

\bibitem[Milone et al.(2012)]{milone2012} Milone, A.~P., Piotto, G., Bedin, L.~R., et al.\ 2012, \apj, 744, 58 

\bibitem[Milone et al.(2010)]{milone2010} Milone, A.~P., Piotto, G., King, I.~R., et al.\ 2010, \apj, 709, 1183 

\bibitem[Norris \& Freeman(1979)]{norris1979} Norris, J., \& Freeman, K.~C.\ 1979, \apjl, 230, L179

\bibitem[Norris et al.(1981)]{norris1981} Norris, J., Cottrell, P.~L., Freeman, K.~C., \& Da Costa, G.~S.\ 1981, \apj, 244, 205 

\bibitem[Norris \& Smith(1985)]{norris1985} Norris, J., \& Smith, G.~H.\ 1985, \aj, 90, 2526 


\bibitem[Origlia et al.(2006)]{origlia2006} Origlia, L., Valenti, E., Rich, R.~M., \& Ferraro, F.~R.\ 2006, \apj, 646, 499 

\bibitem[Osborn(1971)]{osborn1971} Osborn, W.\ 1971, The Observatory, 91, 223


\bibitem[Pancino et al.(2010a)]{pancino2010a} Pancino, E., Carrera, R., Rossetti, E., \& Gallart, C.\ 2010a, \aap, 511, A56 

\bibitem[Pancino et al.(2010b)]{pancino2010b} Pancino, E., Rejkuba, M., Zoccali, M., \& Carrera, R.\ 2010b, \aap, 524, A44 

\bibitem[Pietrinferni et al.(2004)]{pietrinferni2004} Pietrinferni, A., 
Cassisi, S., Salaris, M., \& Castelli, F.\ 2004, \apj, 612, 168 

\bibitem[Piotto et al.(2007)]{piotto2007} Piotto, G., Bedin, 
L.~R., Anderson, J., et al.\ 2007, \apjl, 661, L53

\bibitem[Platais et al.(2011)]{platais2011} Platais, I., Cudworth, K.~M., Kozhurina-Platais, V., et al.\ 2011, \apjl, 733, L1 

\bibitem[Renzini(2008)]{renzini2008} Renzini, A.\ 2008, \mnras, 391, 354 

\bibitem[Smiljanic et al.(2009)]{smiljanic2009} Smiljanic, R., Gauderon, R., North, P., et al.\ 2009, \aap, 502, 267 

\bibitem[Smolinski et al.(2011a)]{smolinski2011a} Smolinski, J.~P., 
Lee, Y.~S., Beers, T.~C., et al.\ 2011a, \aj, 141, 89 

\bibitem[Smolinski et al.(2011b)]{smolinski2011} Smolinski, J.~P., Martell, S.~L., Beers, T.~C., \& Lee, Y.~S.\ 2011b, \aj, 142, 126 

\bibitem[Stetson(1987)]{stetson1987} Stetson, P.~B.\ 1987, \pasp, 99, 191 

\bibitem[Stetson(1994)]{stetson1994} Stetson, P.~B.\ 1994, \pasp, 106, 250 

\bibitem[Stetson et al.(2003)]{stetson2003} Stetson, P.~B., Bruntt, H., \& Grundahl, F.\ 2003, \pasp, 115, 413

\bibitem[Stoughton et al.(2002)]{stoughton2002} Stoughton, C., Lupton, R.~H., Bernardi, M., et al.\ 2002, \aj, 123, 485 

\bibitem[Tucker et al.(2006)]{tucker2006} Tucker, D.~L., Kent, S., Richmond, M.~W., et al.\ 2006, Astronomische Nachrichten, 327, 821 

\bibitem[Twarog et al.(2011)]{twarog2011} Twarog, B.~A., Carraro, 
G., \& Anthony-Twarog, B.~J.\ 2011, \apjl, 727, L7 

\bibitem[Yanny et al.(2009)]{yanny2009} Yanny, B., Rockosi, C., 
Newberg, H.~J., et al.\ 2009, \aj, 137, 4377 
\end{thebibliography}
\end{document}